\documentclass[journal]{IEEEtran}
\usepackage[caption=false,font=footnotesize]{subfig}
\usepackage{graphicx}
\graphicspath{{grafiken/}}
\begin{document}
\title{Development of High-Resolution Muon Tracking\\
Systems Based on Micropattern Detectors}

\author{Jonathan~Bortfeldt, %~\IEEEmembership{LMU~Munich,}
        Otmar~Biebel, %~\IEEEmembership{LMU~Munich,}
        David Heereman %~\IEEEmembership{IIHE,~ULB-VUB,~Brussels,}
        and~Ralf~Hertenberger,~\IEEEmembership{LMU~Munich}% <-this % stops a space
\thanks{Manuscript received May 20, 2011. This work was supported by the DFG cluster of excellence on ``Origin and Structure of the Universe'' and by the Maier-Leibnitz-Laboratory Garching.}% <-this % stops a space
\thanks{J.~Bortfeldt, O.~Biebel and R.~Hertenberger are with LS Schaile, Department of Physics, Ludwig-Maximilians-Universit{\"a}t M{\"u}nchen, Germany (telephone: +49-89-289-14142, e-mail: jonathan.bortfeldt@cern.ch).}%
\thanks{D. Heereman, was with LS Schaile, Ludwig-Maximilians-Universit{\"a}t M{\"u}nchen, Germany. He is now with the Interuniversity Institute for High Energies (ULB-VUB), Brussels, Belgium.}%
}
\maketitle
\thispagestyle{empty}

\begin{abstract}
A muon tracking system consisting of four 9\,cm$\times$10\,cm sized bulk Micromegas detectors with 128\,$\mu$m amplification-gap and two 10\,cm$\times$10\,cm triple GEM detectors is foreseen for high-precision tracking of 140\,GeV muons at the H8 beamline at CERN with a rate of up to 10\,kHz and an overall resolution below 40\,$\mu$m. Larger detectors with an active area of 0.5\,m$^2$ and more are under development for detector studies in high neutron or gamma ray background environments at the Gamma Irradiation Facility at CERN and the Munich tandem accelerator.
Signal studies of both detector types have been performed by recording cosmic muon and 5.9\,keV X-ray signals with a single charge-sensitive preamplifier using several gas-mixtures of Ar:CO$_2$. The signals were digitized using 1\,GHz VME based flashADCs with 2520 sampling points. The analysis of the complete signal-cycles allows for the determination of rise times, pulse heights, timing fluctuations and discrimination of background, resulting in a FWHM energy resolution of about 20\% and detection efficiencies of 99\% and more. Models for signal formation in both detector types will be presented.
The single detector spatial resolution of $80\mu$m was measured using a fast Gassiplex based strip readout with readout strips of 150\,$\mu$m width and a pitch of 250\,$\mu$m. The Gassiplex readout, formerly used at the HERMES experiment, had to be substantially adapted. No more crosstalk or non-linearities were observed after reconfiguration of the multiplexing amplifier on the frontend boards. The observed spatial resolution is limited by multiple scattering of the cosmic muons used in the laboratory.
We also report on the sensitivity to gamma- and neutron background and on the behaviour of spatial resolution as a function of background rates.
\end{abstract}

\section{Introduction}
\IEEEPARstart{M}{icropattern} gaseous detectors are intrinsically high rate capable. Short driftpaths for ions from gas amplification are the key for reducing space charge effects, common in wire chambers, and for low occupancy. The amplification region in Micromegas (MICROMEsh GAS detector) \cite{giomataris:mm}, is formed by the readout anode structure and a fine microgrid. The micromesh is held at a distance of 128\,$\mu$m to the anode by photolithographically etched pillars. Gas Electron Multipliers (GEM) \cite{sauli:gem}, consist of a thin metal-clad insulating polymer foil into which conical holes have been etched. Upon application of a moderate voltage to the metal layers, electrons, drifting into the holes, are amplified efficiently. Micromegas and GEM detectors have been successfully used during the past years in the COMPASS experiment \cite{bernet:compassmm}, \cite{ketzer:compassgem} at CERN as well as in several other applications \cite{sauli:gemapplication}, \cite{kousouris:castmm}. Furthermore, large size Micromegas development is ongoing for the planned high luminosity upgrade of the ATLAS detector at CERN \cite{alexopoulos:mamma}.

The bulk Micromegas were built at the CERN PCB workshop and assembled with slight modifications at the LMU Munich \cite{jona:diplarbeit}. A woven stainless steel mesh with a wire diameter of 18\,$\mu$m and a pitch of 45\,$\mu$m forms the 128\,$\mu$m wide amplification region with the anode structure, consisting of 360 strips with 100\,mm length, 150\,$\mu$m width and 250\,$\mu$m pitch, resulting in an active readout area of 100\,mm$\times$90\,mm. Electrons, produced by ionization in the 5\,mm to 7\,mm wide drift region, formed by the cathode and the mesh, drift towards the amplification region in a homogeneous 1.5\,kV/cm electric field. Avalanche amplification, caused by the electric field of about 41\,kV/cm in the amplification region, results in a gas gain of about 2500 for a gas mixture Ar:CO2 93:7 at NTP.

We developed triple GEM detectors \cite{david:diplarbeit}, based on framed CERN GEM foils with an active area of 100\,mm$\times$100\,mm. The GEM foils consist of 50\,$\mu$m thick Kapton, double-sidedly coated with 5\,$\mu$m copper. At a mutual pitch of 140\,$\mu$m, conical holes with an outer diameter of 70\,$\mu$m and an inner diameter of 50\,$\mu$m are etched into the foil. 
Negative charges, produced in the region between cathode and upper GEM foil, drift towards the GEM in a 1.5\,kV/cm electric field. A potential difference of around 350\,V between two copper layers of each GEM cause high electric fields on the order of 70\,kV/cm in the holes such that gas amplification occurs. Gradually rising fields between the GEM foils extract part of the produced negative charge, such that sufficiently high signals are detectable on the readout structure. Several prototypes have been built, the latest consists of a cathode of aluminized mylar held at a distance of 3\,mm above the stack of three GEM foils, which are separated by 2\,mm from each other and from the readout structure, formed by 360 strips with 100\,mm length, 150\,$\mu$m width and 250\,$\mu$m pitch. We will refer to the region between cathode and upper GEM foil as drift region, the gaps between the foils are called transfer gaps and the region between anode and lower GEM foil is referred to as induction gap.

\section{Characteristics of the Micromegas and GEM detectors}
\subsection{Flash ADC Setup}
In order to conduct signal studies and investigate the general behavior of both detector types, signals of a single charge sensitive preamplifier were recorded, using a 1\,GHz Flash ADC\footnote{VME based CAEN 12 bit FADC V1729}. A typical signal is shown in Fig.\,\ref{fig:measchaanode}. Different groups of anode readout strips were connected, forming single readout planes of different area. All other strips were grounded. The edge of the charge signals was parametrized by a fermi-like function. Pulse heights, rise times and timing fluctuation, provided that the readout was triggered externally, can be deduced from this fit function. 

For measurements with a $^{55}$Fe source, a trigger signal was derived from the preamplifier signal by amplification and shaping with a fast timing filter amplifier and subsequent discrimination in a low threshold discriminator\footnote{ORTEC 474 TFA and CAEN N840 discriminator}. The dominant $^{55}$Fe$_{K\alpha}$-line at 5.9\,keV is used for studies of the pulse height dependence on several operational parameters. 

Cosmic signals were triggered by coincident signals from two scintillators above and below the detector and the detector itself. By comparing the number of coincident scintillator triggers to the number of threefold coincidences, the efficiency was calculated. A geometric correction had to be applied to the number of scintillator triggers, since their active area was slightly larger, than the active region of the gas detectors.

\subsection{Gas Mixture- and Pressure Control-System}
In order to obtain stable and reproduceable operation parameters, and prevent oxygen from entering the detectors, the gas pressure was controlled and set to a slight overpressure of 1013\,mbar. Several Brooks Mass Flow Controllers adjust the gas composition, that is mixed in a gas mixer and passed to the detector. A regulation valve in combination with a pressure meter and a PI-controller stabilizes the pressure in the system at the preset value. Usually an overall flow on the order of 1\,ln/h was used.

\subsection{Efficiency and Energy Resolution}
\begin{figure}[!t]
\centerline{\subfloat[GEM]{\includegraphics[width=0.5\columnwidth]{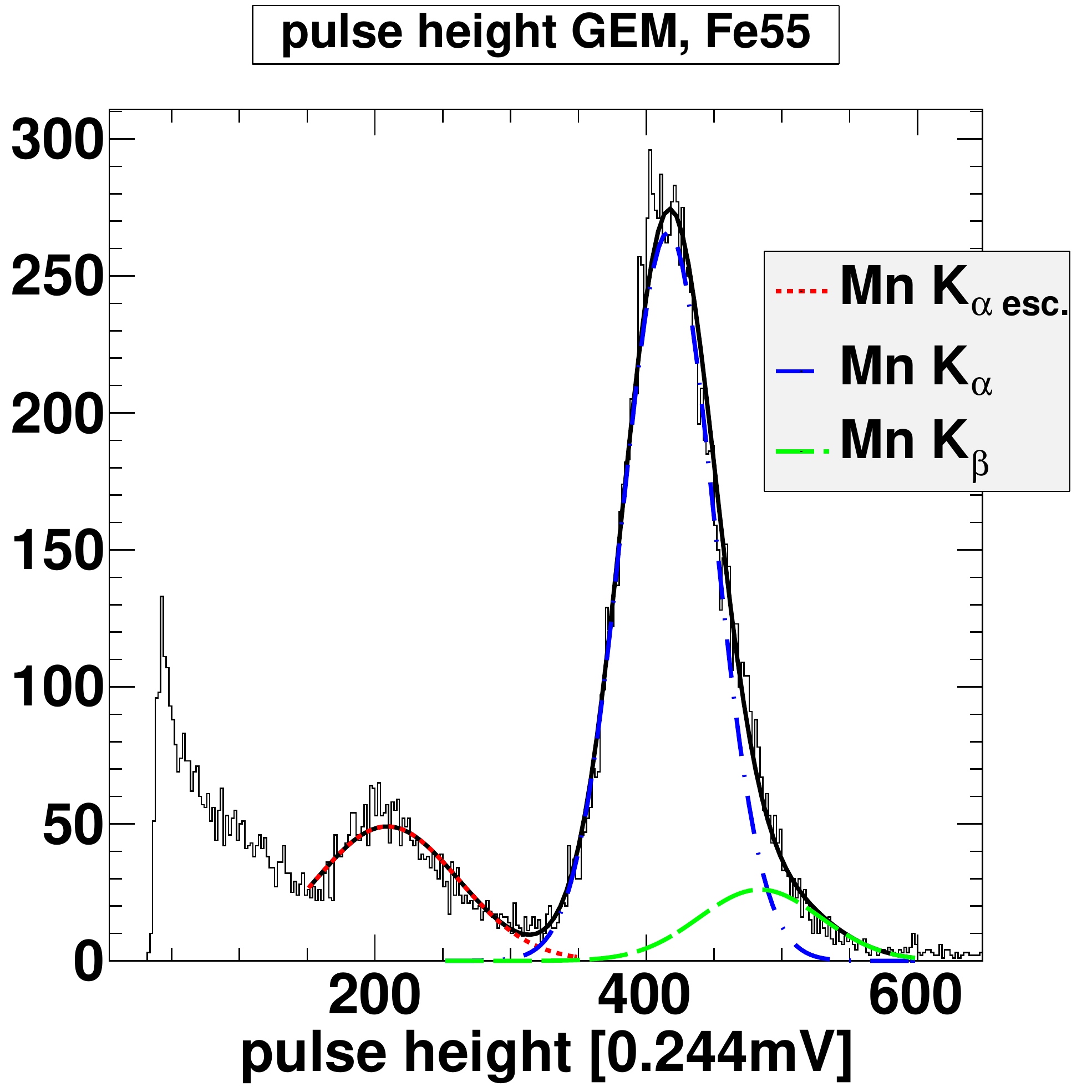}%
\label{fig:pulseheight_gem_190}}
\hfil
\subfloat[Micromegas]{\includegraphics[width=0.5\columnwidth]{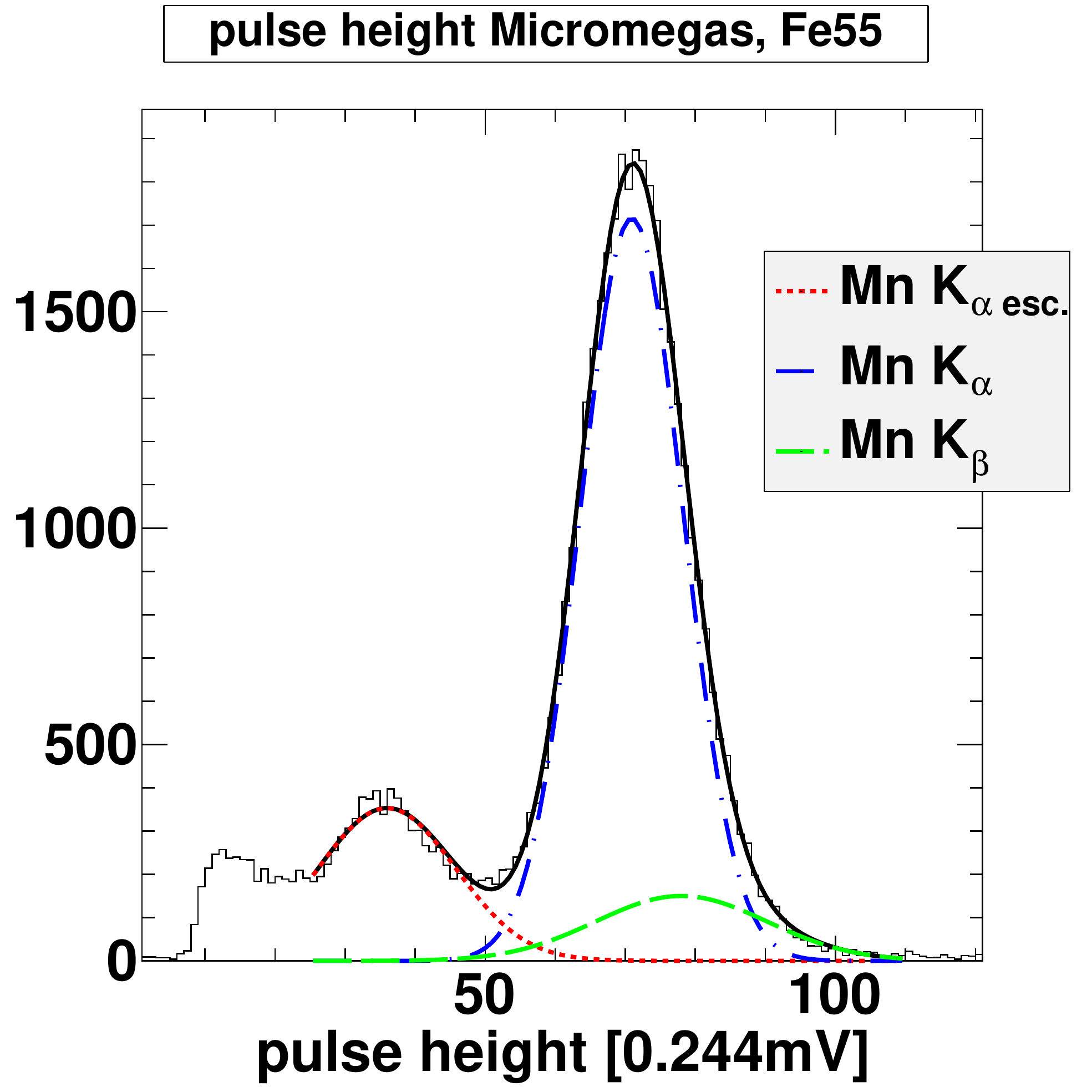}%
\label{fig:pulseheight_mm_190}}}
\caption{$^{55}$Fe spectra in GEM detectors (\ref{fig:anodesig}) and Micromegas (\ref{fig:meshsig}).}
\label{fig:pulseheight_fe55}
\end{figure}

Detector efficiencies for muons of more than 98\% are achieved in both detector types. In Micromegas, the efficiency is limited by the mesh supporting pillars, covering 1.1\% of the active area. The FWHM energy resolution for 5.9\,keV X-rays is 19.2\% for the triple GEM detector and 24.3\% for the Micromegas. It has been determined by fitting Gaussian functions to the three lines in the $^{55}$Fe spectra, as shown in Fig.\,\ref{fig:pulseheight_fe55}. The standard deviation of the dominant 5.9\,keV peak is multiplied by 2.355 to obtain the FWHM energy resolution.

\subsection{Signal Formation in Micromegas}
The anode charge signals observed in Micromegas' are the sum of the very fast electron signal and the slower induction signal, created by the drifting and vanishing ion cloud. They can be calculated analytically. Similar results have been stated in \cite{mathieson:avalcountersignal} for a parallel plate avalanche chamber. The measured pulse height depends in a complex way on the capacitances within the detector and the preamplifier. The reaction of this capacitor network to signals is evaluated, using the circuit simulation program LT Spice IV \cite{ltspice:usersguide}.

\begin{figure}[!t]
	\centering
		\includegraphics[width=0.5\columnwidth]{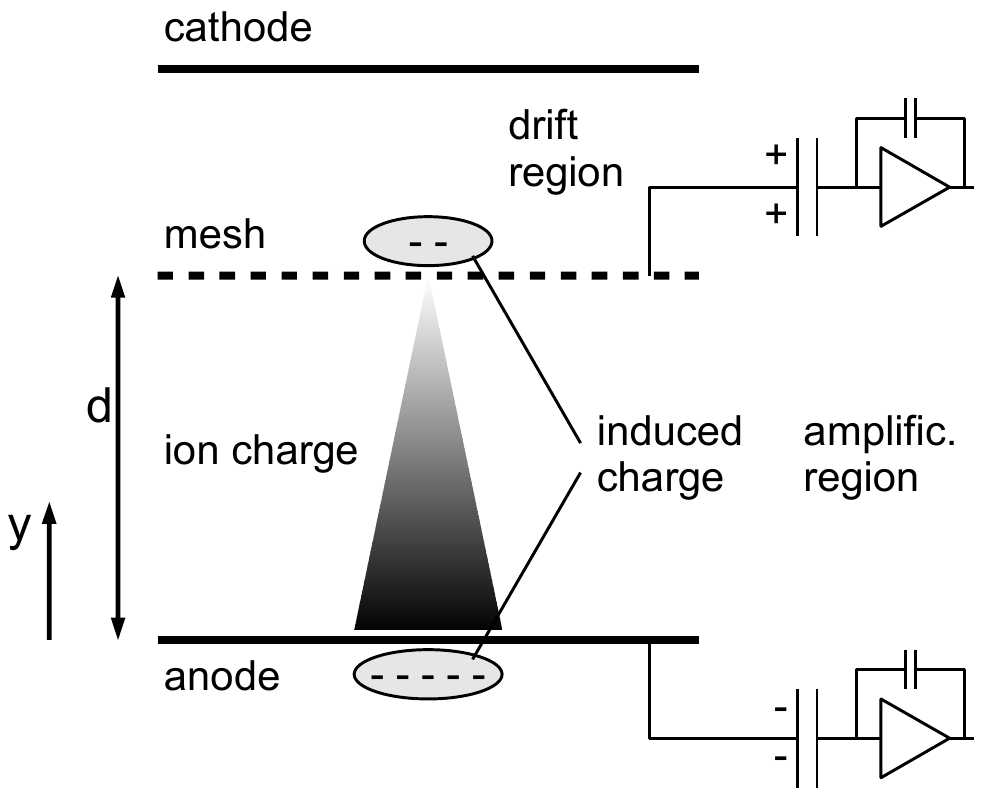}
		\caption{Schematic, showing the situation after completed gas amplification at t=0. The distance cathode-mesh (5-7\,mm) and mesh-anode (128$\,\mu$m) are not drawn to scale.}
		\label{fig:MeshAnodeT0}
\end{figure}

A charge of $-q_0=-en$ entering the amplification region with width $d$ creates equal amounts of positive and negative charge through gas amplification. Since the mobility of electrons is about three orders of magnitude larger than that of ions, the temporal duration of the process of gas amplification can be neglected. We assume the gas amplification to be instantaneous. Gas amplification, according to Townsend gives the electron charge $Q_e(y)$ as well as the ion charge $Q_i(y)$ as a function of the distance $y$ from the anode
\begin{eqnarray}
Q_{e}(y) =& -q_0 \exp(\alpha(d-y))\\\label{eq:ioncharge}
Q_{i}(y) =& q_0 \exp(\alpha(d-y)) - q_0~,
%\label{eq:}
\end{eqnarray}

where $\alpha$ is the first Townsend coefficient. At $t=0$ the situation is as follows: All produced electron charge $q_e=-q_0\exp(\alpha d)$ has reached the anode, the positive ion cloud, described by the charge density in planes with thickness $\mathrm{d}y$ parallel to anode and mesh $\mathrm{d}Q_i/\mathrm{d}y=\rho_i(y,t=0) = \alpha q_0\exp(\alpha(d-y))$, induces a negative surface charge $q_{ai}$ on the anode and $q_{mi}$ on the mesh. No ion charge has reached the mesh yet, i.e.~$q_i(t=0)=0$. It should be noted, that we will calculate the actually measureable signal. That implies, that a negative surface charge on anode or mesh leads to a positive detectable charge, explaining the ``$-$'' in (\ref{eq:qanode}) and (\ref{eq:qmesh}). The detectable charges on anode $q_a$ and mesh $q_m$ are then given by
\begin{eqnarray}\label{eq:qanode}
q_a =& \underbrace{q_e}_{< 0} - \underbrace{q_{ai}}_{< 0}\\
\label{eq:qmesh}
q_m =& \underbrace{q_i}_{> 0} - \underbrace{q_{mi}}_{< 0}~.
\end{eqnarray}

\begin{figure}[!t]
\centerline{\subfloat[anode signal]{\includegraphics[width=0.5\columnwidth]{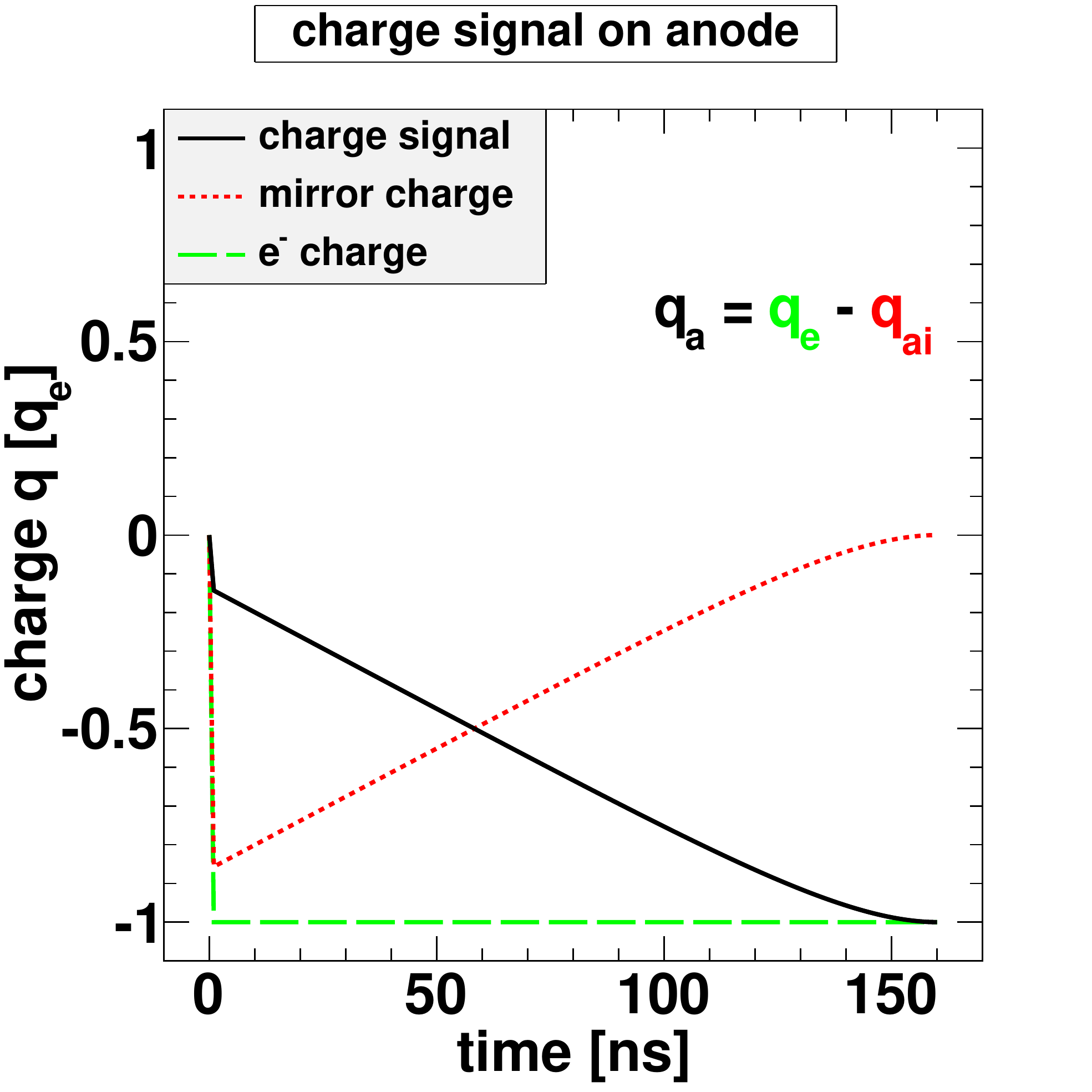}%
\label{fig:anodesig}}
\hfil
\subfloat[mesh signal]{\includegraphics[width=0.5\columnwidth]{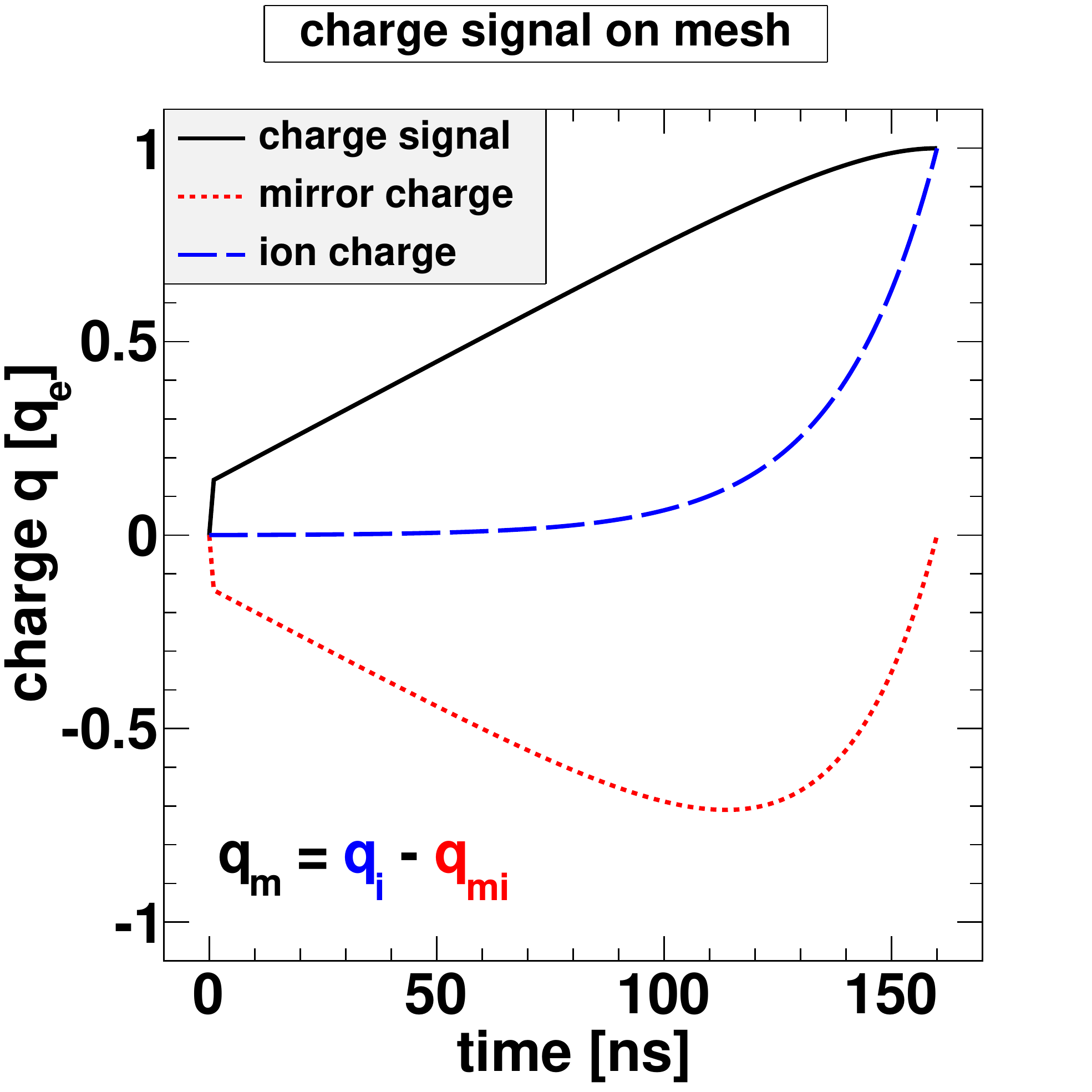}%
\label{fig:meshsig}}}
\caption{Calculated observable charge signals for anode (\ref{fig:anodesig}) and mesh (\ref{fig:meshsig}).}
\label{fig:calcsignals}
\end{figure}

Taking into account the constant ion drift velocity $v$ in the homogeneous amplification field, the observable signals can be calculated. The time resolved signals on mesh and anode are displayed in Fig.\,\ref{fig:calcsignals} \cite{jona:sigform}.

\begin{figure}[!t]%
\centering
\includegraphics[width=0.5\columnwidth]{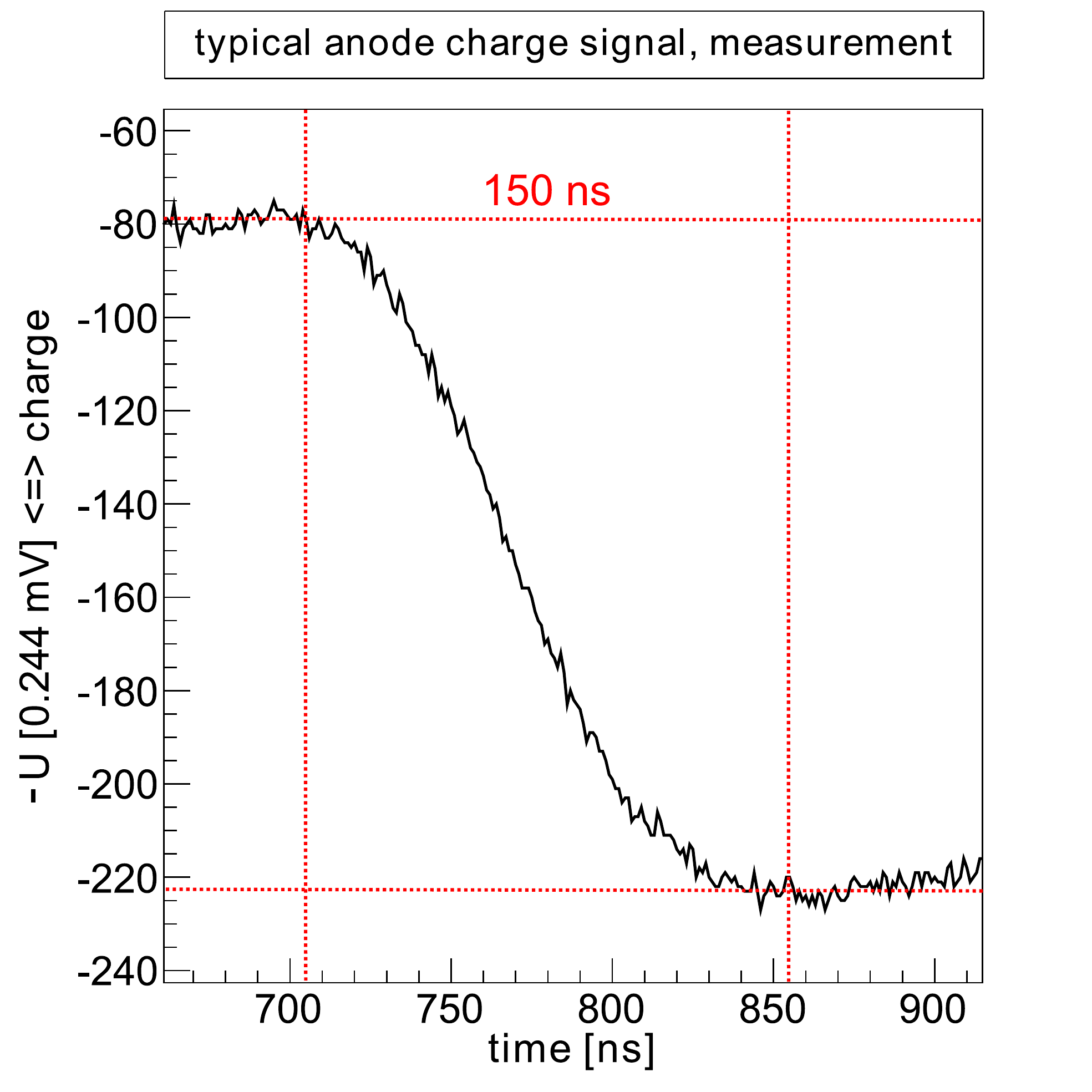}%
\caption{Typical charge signal (inverted).}%
\label{fig:measchaanode}%
\end{figure}

A measured anode charge signal is shown in Fig.\,\ref{fig:measchaanode}. Its amplitude rises to its maximum value within 150\,ns and shows linear rise behavior, smoothed out at the edges by the preamplifier. This behavior is reproduced by the analytic calculation, although the applied preamplifier, showing an intrinsic rise time of 25\,ns, is not fast enough to resolve the steep initial rise of the signal. Nevertheless, with a fast current sensitive preamplifier and a small detector, this fast and very short signal, formed by 14\% of the total signal charge, has been observed (\cite{charpak:mmmultipurpose}) and can be used for sub nanosecond triggering purposes.

\begin{figure}[!t]
\centerline{\subfloat[]{\includegraphics[width=0.5\columnwidth]{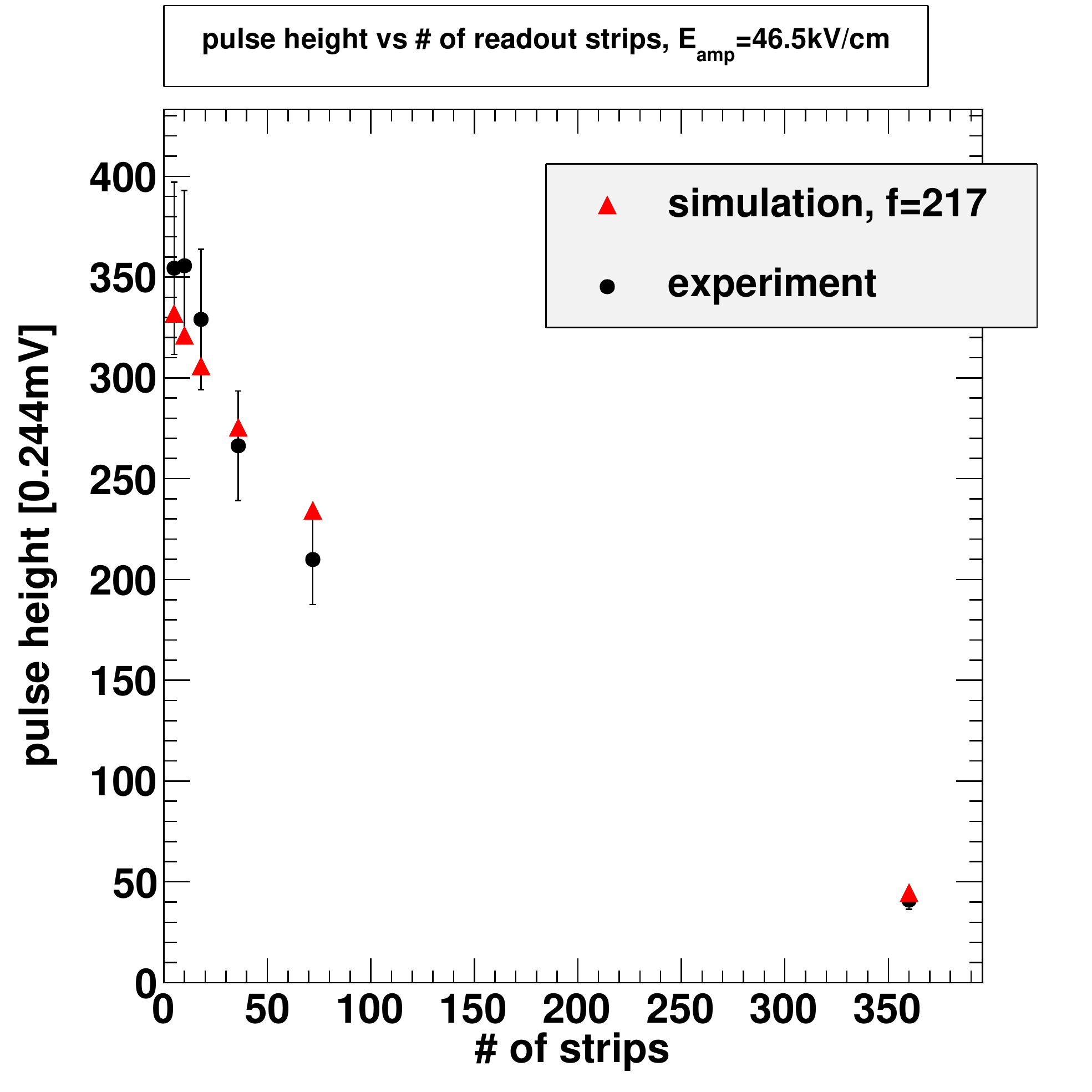}%
\label{fig:1111PhVsNoStrips}}
\hfil
\subfloat[]{\includegraphics[width=0.5\columnwidth]{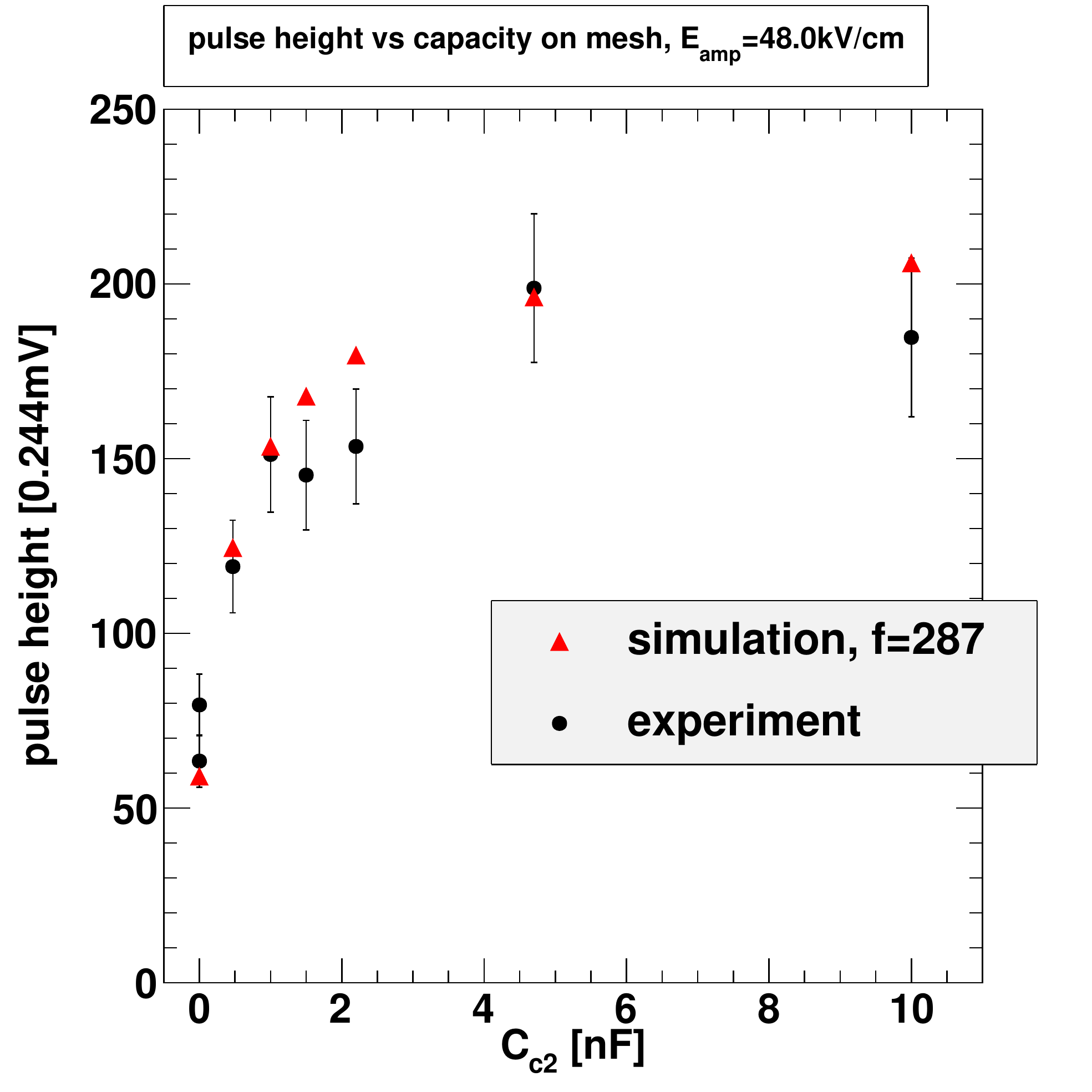}%
\label{fig:1111PheiCsCap}}}
\caption{Pulse height as a function of the number of connected readout strips (\ref{fig:1111PhVsNoStrips}) and as a function of the mesh to ground capacitance (\ref{fig:1111PheiCsCap}). Black circles represent data, red triangles calculated expectation.}
\label{fig:PheiVsCaps}
\end{figure}

The pulse height of signals depends strongly on the complex capacitor network, formed by the anode readout strips, the mesh and detector ground. Two seperate studies have been conducted, using the $^{55}$Fe$_{K\alpha}$-line at 5.9\,keV. Interconnecting different numbers, $n$, of neighboring readout strips to form a single readout plane and grounding the other $360-n$ strips, revealed an increase of the pulse height by almost a factor of 10 when going from 360 to 18 interconnected readout strips (Fig.\,\ref{fig:1111PhVsNoStrips}). Varying the mesh to ground capacitance by adding discrete capacitors, while reading out all 360 strips connected together, caused an increase of the pulse height by a factor of 4 (Fig.\,\ref{fig:1111PheiCsCap}).

\begin{figure}[!t]%
\includegraphics[width=\columnwidth]{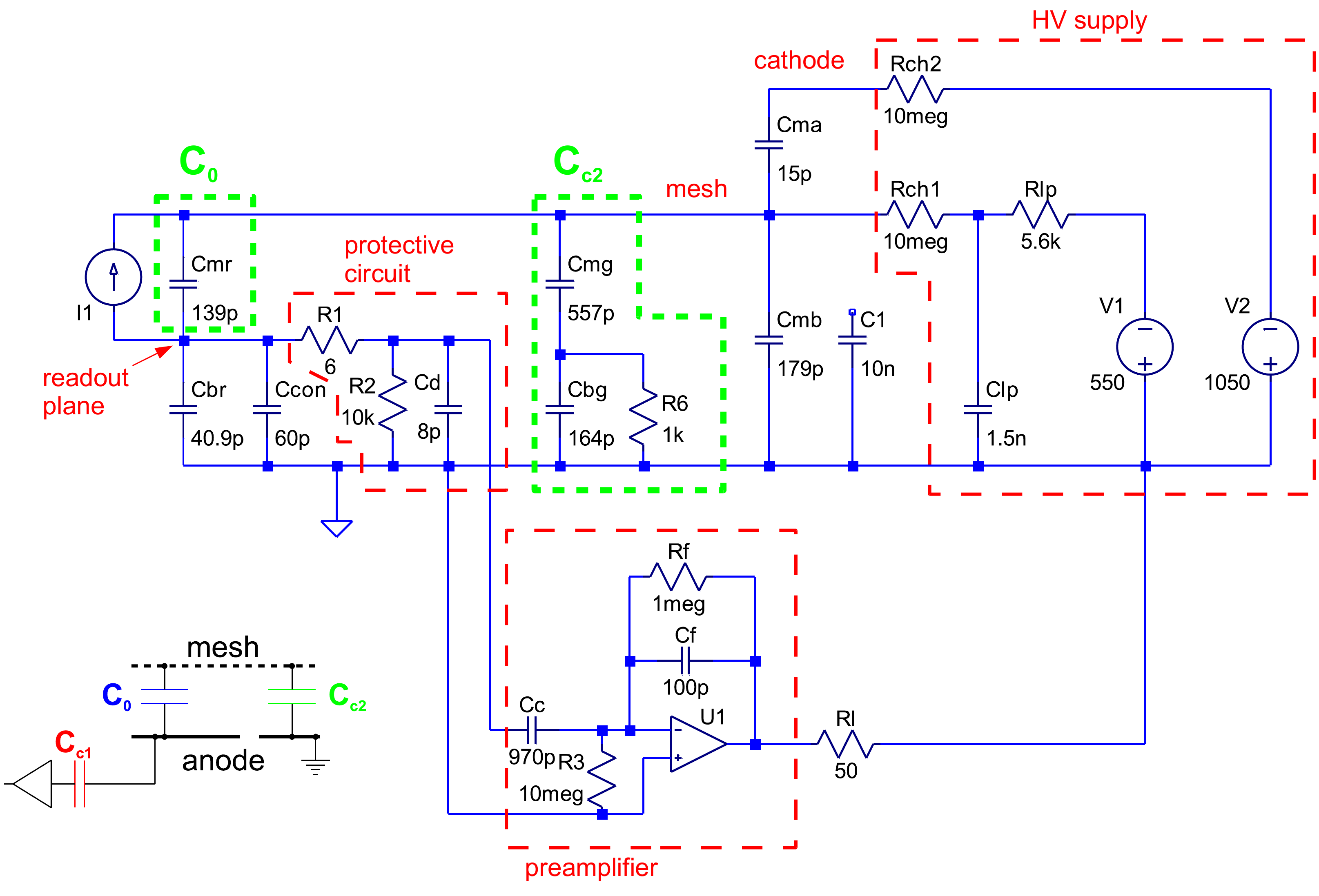}%
\caption{Capacitor network, formed by the detector and the readout electronics.}%
\label{fig:MicomSimSchRev2}%
\end{figure}

Both effects can be quantitatively described by the parameter-free capacitor model, shown in Fig.\,\ref{fig:MicomSimSchRev2}, that allows for the numerical calculation of signals, using the circuit simulation program LTSpice IV. The agreement of the data and the numerical calculation can be seen in Fig.\,\ref{fig:PheiVsCaps}.

\subsection{Signal Formation in GEM Detectors}
\begin{figure}[!t]%
\centering
\includegraphics[width=0.6\columnwidth]{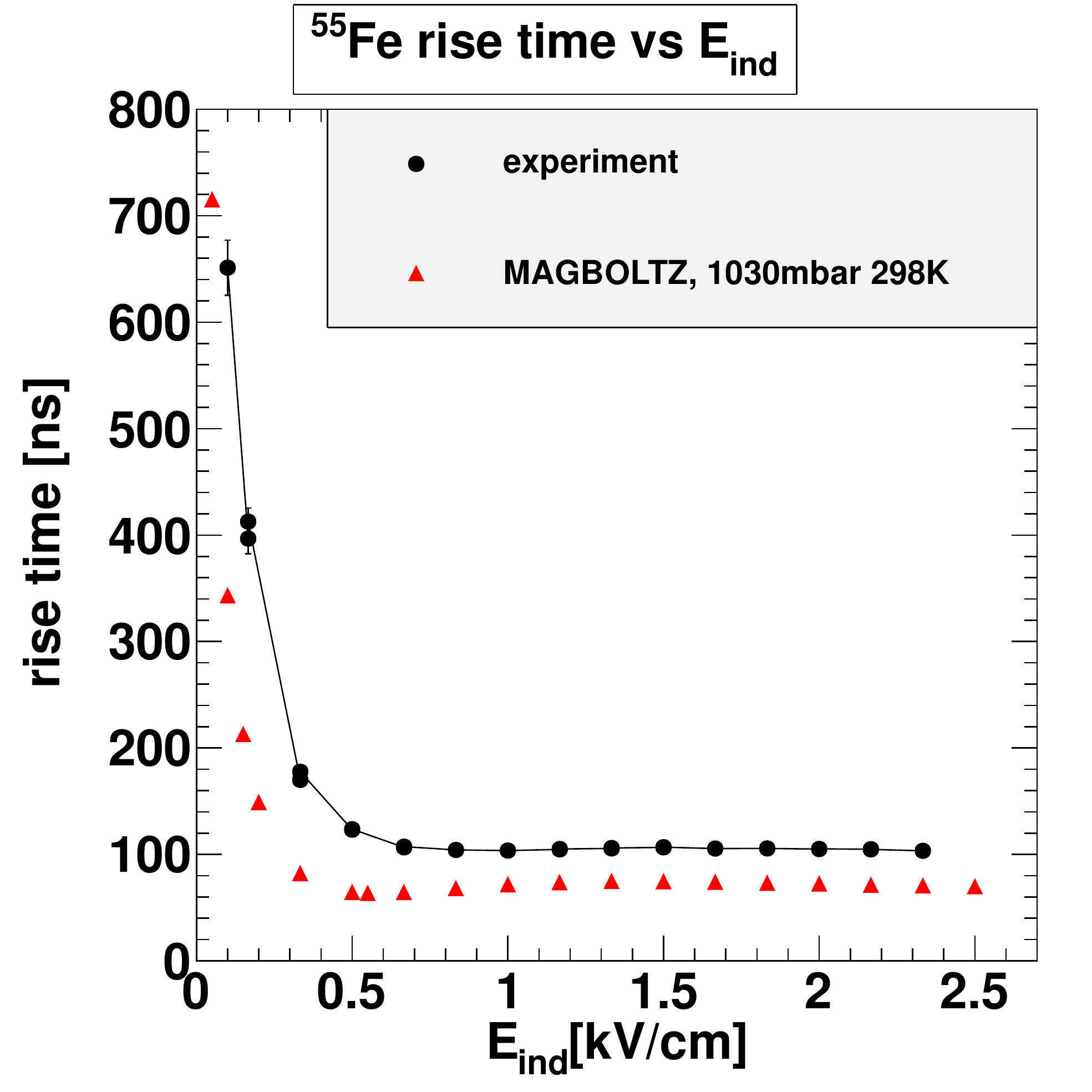}%
\caption{Rise time of $^{55}$Fe signals as a function of the induction field (black circles) and computed electron drift times in the induction gap (red triangles).}%
\label{fig:051011GEMTRiseVsEind}%
\end{figure}

The temporal behavior of the signal and the influence of the induction and drift gap parameters on the signal in double GEM detectors has been treated in \cite{guedes:gemsignal}. Signal formation in GEM detectors is less complicated, since the influence of the ion drift on the anode signal can be neglected. The observed charge signal is created by induction of the drifting electron cloud towards the anode and shows a linear fall under the assumption of a constant electron drift velocity,  
\begin{equation}
q_a(t) = -\frac{q_0 G_{\mathrm{eff}} v_{\mathrm{drift},e}}{d} t~, 
\end{equation}
where $G_{\mathrm{eff}}$ is the effective gas gain and $v_{\mathrm{drift},e}$ the induction field dependent electron drift velocity. The rise time of charge signals, produced by a localized amount of charge, is given by the electron drift time in the induction region. In Fig.\,\ref{fig:051011GEMTRiseVsEind} the measured 10\%-90\% signal rise time is plotted as a function of the induction field together with the calculated drift time of the electrons in the induction gap. The measured rise times are systematically larger than the simulated electron drift times, but the functional behavior is correctly reproduced. Part of this discrepancy might be due to the non-neglegible rise time of the preamplifier, differing GEM-anode distances or gas parameters. As described in the previous section for Micromegas, we observe a significant rise of the pulse height of 5.9\,keV X-ray signals in the GEM detector with a decreasing number of interconnected readout strips, although the increase is not as large as with Micromegas.

\section{Tracking System}
\subsection{Setup}
\begin{figure}[!t]%
\includegraphics[width=\columnwidth]{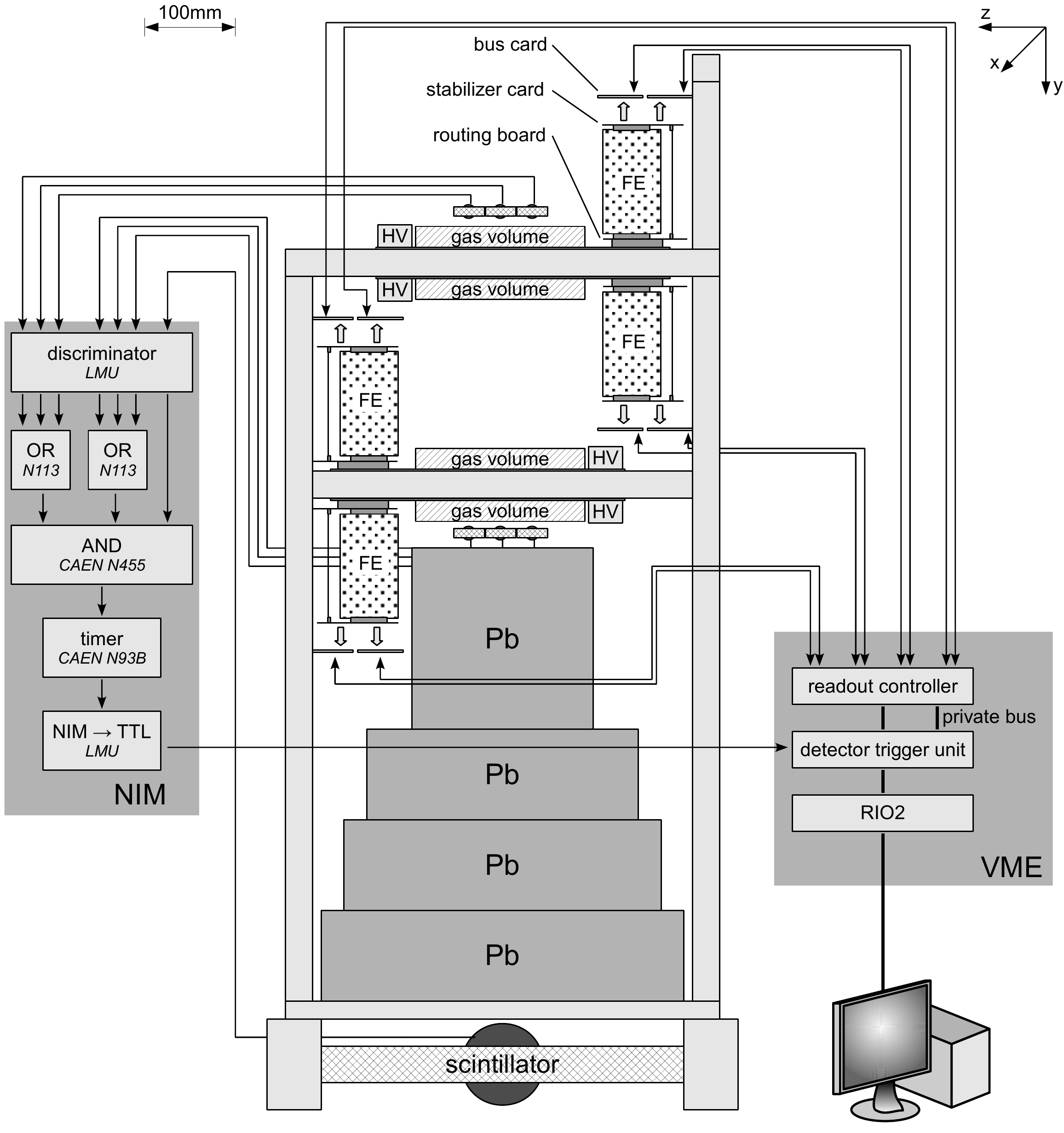}%
\caption{Setup and trigger logic of the cosmic muon tracking system. The gas volume of the four Micromegas detectors is displayed as hatched areas, the trigger hodoscope and the scintillator below the lead absorbers are drawn cross-hatched. Details regarding the Gassiplex based readout system (FEs are dotted)are given in \cite{jona:diplarbeit}.}%
\label{fig:GassiSetup4Micoms}%
\end{figure}

The cosmic muon tracking system consists of four Micromegas, mounted on a stable aluminum frame (Fig.\,\ref{fig:GassiSetup4Micoms}). It is used for determining the spatial resolution of Micromegas in different radiative background environments. All readout strips point into the $z$-direction, such that two-dimensional track reconstruction in the $x$-$y$-plane is possible. Each detector is read out by six frontend boards with 64 channels each, originally developed for the HADES-RICH detector and adapted for Micromegas like signals \cite{kastenmueller:hadesro}. On each frontend board, charge to voltage conversion is achieved by four Gassiplex chips with 16 channels each. Digitization, discrimination and multi event-buffering is performed on the frontend boards, directly mounted on the detector. Unlike in the HERMES scintillating fibre detector, that used the same boards with a slightly different hardware configuration, we observed neither crosstalk nor signal non-linearities. This was achieved by reconfiguring the onboard multiplexing amplifier, that is used for matching the Gassiplex output to the input range of the ADC. An event readout rate of about 10\,kHz is expected to be realizable.

A three-fold segmented 100\,mm$\times$100\,mm scintillator trigger hodoscope provides information about the third track coordinate. Triggering muons have an energy larger than 600\,MeV, if they can cross the 450\,mm thick lead absorber above a 400\,mm$\times$800\,mm large scintillator panel. The hodoscope data are acquired via unused channels on a frontend in the uppermost Micromegas, such that a precise offline selection of events is possible. Due to the relatively small solid angle, the trigger rate is $(3.6\pm0.3)$\,min$^{-1}$. Stable operation over weeks has been observed.

\subsection{Data Analysis}
Primary hits in each detector are found by using the center of gravity method for the strips around the one with maximum pulse height (94\% of tracks). Only strips are regarded, that have passed a criterion for the pulse height, beeing 3$\sigma$ above background. In case of hit ambiguities in one detector, those hits are regarded as true, that lie on a 5.25\,mm wide road, defined by hits in the other three detectors (6\% of tracks). The single detector spatial resolution is determined as follows: A track, defined by hits $r_1$ and $r_2$ in two detectors is interpolated into a third detector, sandwiched by them. The difference between the actual hit in the third detector $r_3$ and the prediction through interpolation is $\delta = r_3 - r_2 d_{13}/d_{12} - r_1(1-d_{13}/d_{12})$, where $d_{12}$ is the distance between the 1st and 2nd detector. For each four-detector-track, four different combinations of three detectors are possible, yielding four $\delta$s. Using Gaussian error propagation, a set of four equations can be derived, connecting the widths of the four $\delta$ distributions $\Delta \delta\in [100\,\mu\mathrm{m},130\,\mu\mathrm{m}]$ with the spatial resolutions in all detectors $\Delta r_i$ with $1\leq i \leq 4$. We assume, that the resolution is $\Delta r_i = \sqrt{\Delta r_{\mathrm{int}}^2 + \Delta r_{\mathrm{ms},i}^2}$, where $\Delta r_{\mathrm{int}}$ is the intrinsic spatial resolution, which is equal for all detectors, and $\Delta r_{\mathrm{ms},i}$ is the detector dependent contribution from multiple scattering. Assuming a simple linear dependence for the multiple scattering $\Delta r_{\mathrm{ms},i} = 0.13\,\mu\mathrm{m}/\mathrm{mm}\times d_{ui}$, where $d_{ui}$ is the distance between the uppermost and the $i$-th detector, the intrinsic resolution $\Delta r_{\mathrm{int}}$ is obtained from solving the equation system.

\subsection{Detector Performance in Radiative Background}

We studied the behavior of the tracking system under irradiation with a $^{137}$Cs-source, dominantly emitting gammas with $E_\gamma = 662$\,keV, and a $^{252}$Cf-source, emitting neutrons and gammas with a broad energy distribution with energy up to $E_{n} = 8$\,MeV and $E_{\gamma} = 4$\,MeV,  $\overline{E}_n = 2.1$\,MeV and $\overline{E}_\gamma = 870\,$keV. The sources were positioned right on top of the third Micromegas from the top. Using the $^{137}$Cs-source, this Micromegas registered a hit rate of $(4.2\pm0.2)$\,kHz, corresponding to a  gamma flux rate of $(1.59\pm0.04)\times10^6$\,Hz with an efficiency of $\varepsilon_{\mathrm{Cs}137} = (2.64\pm0.20)\times10^{-3}$ . With the $^{252}$Cf-source, an overall hit rate of $(195\pm10)$\,Hz was measured. Spontaneous fission causes a neutron flux rate in the detector of $(9.1\pm0.3)$\,kHz and a gamma flux rate of $(16.4\pm1.3)$\,kHz. For the efficiency to neutrons, an upper bound of $\varepsilon_{\mathrm{Cf}252,n}<6.8\times10^{-4}$ at 95\% CL was determined. The efficiencies have been determined by simply counting signals and correcting for background hits. The deadtime of the detector can be neglected in this case. A pulse shape discriminating scintillator was used to characterize the $^{252}$Cf-source \cite{alex:privatecomm}. 

Without gamma or neutron irradiation, the mean sparking rate of the four Micromegas was $(0.19\pm0.01)$\,min$^{-1}$. It showed a slight increase to $(0.24\pm0.01)$\,min$^{-1}$ under irradiation by the $^{137}$Cs-source, the sparking rate in the dominantly irradiated Micromegas 3 did not show a significant increase. The $^{252}$Cf-source on the other hand, triggered a rise of the sparking rate to $(0.50\pm0.04)$\,min$^{-1}$. For Micromegas 3, it went up to $(0.74\pm0.05)$\,min$^{-1}$. 

\begin{figure}[!t]%
\centering
\includegraphics[width=0.5\columnwidth]{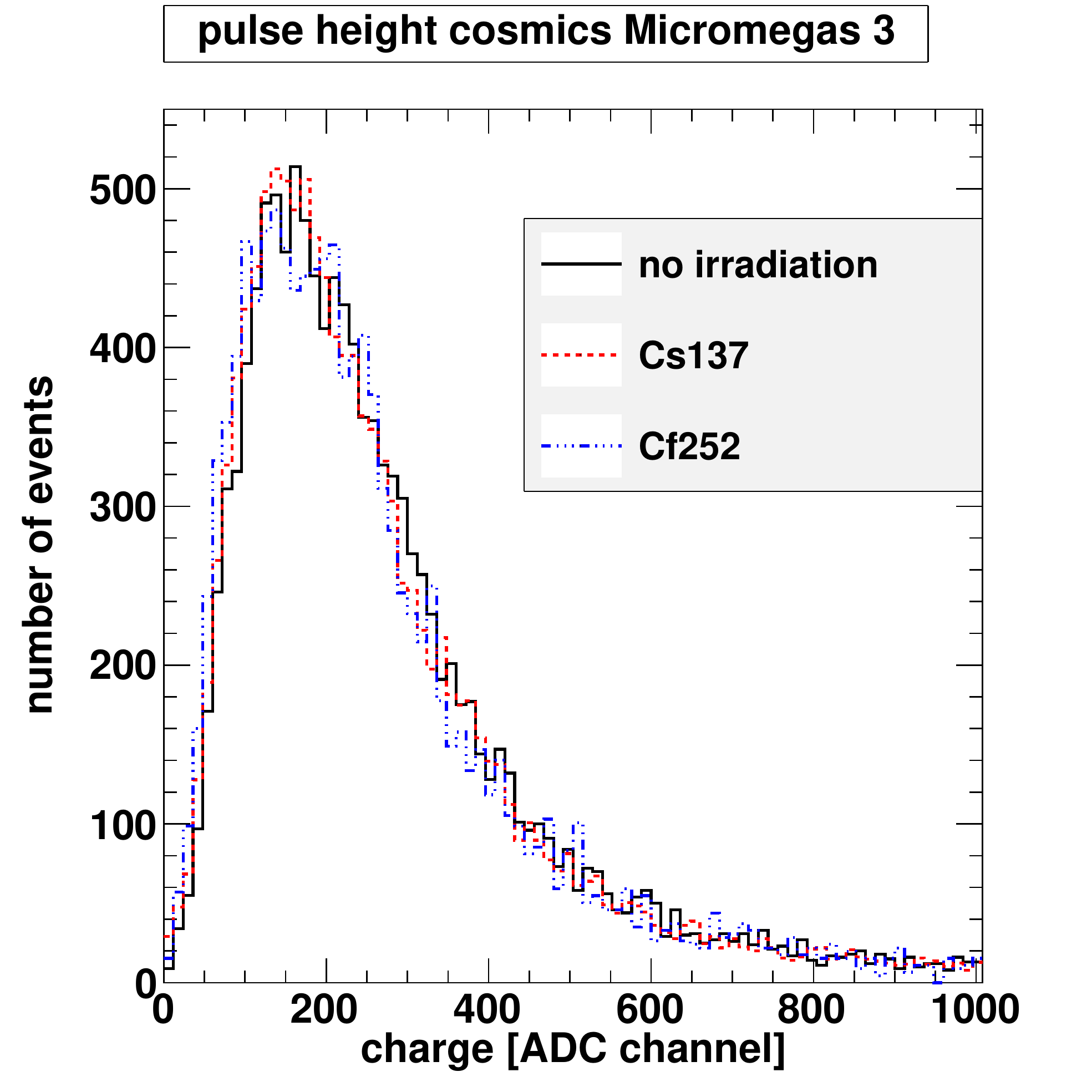}%
\caption{Pulse height spectra for cosmics, representing the total charge in an event, for irradiation with a $^{137}$Cs-source (red, dashed), irradiation with a $^{252}$Cf-source (blue, dash-dotted) and no irradiation (black, solid).}%
\label{fig:051011PHRuns323_322_327}%
\end{figure}

As can be seen from Fig.\,\ref{fig:051011PHRuns323_322_327} where the typical Landau pulse height spectra are displayed, the irradiation did not alter the pulse height of cosmic signals.

The single detector spatial resolution has been determined, using the method, described above. Assuming that the spatial resolution is equal in all detectors, this method yields an intrinsic spatial resolution for cosmic muons of $\Delta r_{\mathrm{int}} = 80\,\mu$m in the run without gamma or neutron background. In measurements with the $^{137}$Cs-source, no degradation of the spatial resolution was visible. The determination of the spatial resolution under neutron irradiation is ongoing, since the pulse height spectra remains unchanged though, no significant degradation is expected.

The spatial resolution of the GEM detector has been determined with a different method and lies in the same range \cite{david:diplarbeit}.

\section{Conclusion}
We developed a muon tracking system consisting of four Micromegas. An upgrade with two triple GEM detectors is forseen. Both detector types have been investigated in depth and models for signal formation have been presented. Efficiencies for muons above 98\% and energy resolutions for 5.9\,keV X-rays of 24\% for the Micromegas and 19\% for the triple GEM have been achieved. The Micromegas spatial resolution has been determined to be 80\,$\mu$m, using cosmic muons. No degradation under irradiation with a $^{137}$Cs-source was visible, for irradiation with $^{252}$Cf, no degradation is expected.
%\newpage

\appendices

\section*{Acknowledgment}
We would like to thank M.~B{\"o}hmer and L.~Maier for providing the Gassiplex readout and J.~Wotschack for the help with and the discussions about Micromegas. Furthermore we would like to thank A.~Ruschke for the characterization of the $^{252}$Cf-source.

\bibliographystyle{IEEEtran}

% that's all folks
\end{document}